
\input harvmac.tex

\def\frac#1#2{{#1\over#2}}

\def\exp{{\rm exp}}

\def\cth{\rm coth}

\mathchardef\ka="101A

\catcode`\@=11
\def\slash#1{\mathord{\mathpalette\c@ncel{#1}}}
\overfullrule=0pt
\def\stackreb#1#2{\mathrel{\mathop{#2}\limits_{#1}}}
\def\steepslash{\c@ncel}
\def\frac#1#2{{#1\over #2}}

\def\inbar{\,\vrule height1.5ex width.4pt depth0pt}
\def\IB{\relax{\rm I\kern-.18em B}}
\def\IC{\relax\hbox{$\inbar\kern-.3em{\rm C}$}}
\def\IP{\relax{\rm I\kern-.18em P}}
\def\IR{\relax{\rm I\kern-.18em R}}
\def\IZ{\relax\ifmmode\mathchoice
{\hbox{Z\kern-.4em Z}}{\hbox{Z\kern-.4em Z}}
{\lower.9pt\hbox{Z\kern-.4em Z}}
{\lower1.2pt\hbox{Z\kern-.4em Z}}\else{Z\kern-.4em Z}\fi}

\def\ddx{\partial_\sigma}

\def\ddt{\partial_\theta}
\catcode`\@=12

\Title{\vbox{\baselineskip12pt\hbox{IASSNS-HEP-92/62}
                \hbox{RU-92/37}}} {\vbox{\centerline{Free Field Representation}
\vskip6pt\centerline{for the Classical Limit of}
\vskip6pt\centerline{Quantum Affine Algebra}}}

\centerline{Sergei Lukyanov \footnote{$\dagger$}
{Research supported  by grant DE-FG05-90ER40559 }\footnote{$^*$}
{On leave of absence from L.D.Landau Institute for Theoretical
Physics, Kosygina 2, Moscow, Russia.}}
\bigskip\centerline{Department of Physics and Astronomy}
\centerline{Rutgers University,Piscataway, NJ 08855-049}
\bigskip\centerline{Samson L. Shatashvili \footnote{$\ddagger$}
{Research supported by DOE grant DE-FG02-90ER40542. }
\footnote{$^\#$}{On leave of absence from St. Petersburg
Branch of Mathematical Institute (LOMI), Fontanka 27, St.Petersburg
191011, Russia.}}
\bigskip\centerline{School of Natural Science}
\centerline{Institute for Advanced Study}
\centerline{Olden Lane}
\centerline{Princeton, NJ 08540}
\vskip .5in

Free field representation for the classical limit of
quantum affine algebra
is constructed by simple deformation of the known
expressions from WZW
theory.

\Date{September, 92}

The progress  achieved during the last decade in CFT is
a result of a successful applications of the representation
theory of            infinite dimensional algebras to
physics problems \ref\bpz{A. A. Belavin, A. M. Polyakov and
A. B. Zamolodchikov, Nucl.Phys. B241 (1984) 333.},
\ref\w{E. Witten, Commun. Math. Phys. 92 (1984) 455.},
\ref\kz{V. G. Knizhnik and A. B. Zamolodchikov,
Nucl. Phys. B247 (1984) 83.}. In the class of CFT
where one can speak about classical theory, representation
theory arises via geometric quantization of classical phase space. For
compact WZW model this is probably best understood and it turned out that
classical phase space for chiral part of the theory can be
viewed as the space of mappings of the segment to the group \ref\as
{A. Alekseev and S. Shatashvili, Commun. Math. Phys. 133 (1990) 353.},
\ref\fad{L. D. Faddeev, Comm.Math.Phys., 132 (1990) 131.},
\ref\lf{S. L. Lukyanov and V. A. Fateev, Int. J. Mod.
Phys. A7 (1992) 853; 1325.},
\ref\gaw{K. Gawedzki, Commun. Math. Phys. 139 (1991) 201.},
\ref\god{M. Chu, P. Goddard, I. Halliday, D. Olive,
A. Schwimmer, Phys.Lett.B 266 (1991) 71.}.
In this language the group element becomes the simplest chiral
vertex operator after quantization. The existence of
Darboux variables on phase space leads to free field representation
for vertex operators and symmetry algebra itself
\ref\zam{A. B. Zamolodchikov, unpublished.},\ref\wak{M. Wakimoto,
Commun. Math. Phys. 104 (1986) 605.} and this could be  viewed as
classical explanation of Feigin-Fuchs-Dotsenko-Fateev
integral representation for conformal blocks. The latter practically is
the most powerful technical way of constructing the correlation functions
in Rational CFT.

There are serious indications in the favor of existence of the analogous
structure  in the massive integrable theories (for latest
developments see \ref\sm{F. A. Smirnov, Dynamical symmetries of
massive integrable models, I ; II, RIMS preprints 772 ; 838.},
 \ref\fr{I. B. Frenkel and N. Yu. Reshetikhin, Commun. Math.
Phys. 146 (1992) 1.},
 \ref\japone{O. Foda and T. Miwa, Int. J. Mod. Phys. A7 Suppl.
1A (1992) 279.}, \ref\japtwo{B. Davies, O. Foda, M. Jimbo, T. Miwa
and A.Nakayashiki, Diagonalization of the XXZ
Hamiltonian by vertex operators, RIMS preprint 873 (1992).},
\ref\japthree{M. Jimbo, K. Miki, T. Miwa and A. Nakayashiki,
Correlation Function of the XXZ model for $\Delta < -1$,
RIMS preprint 877 (1992).}).
 F. Smirnov \sm\ have pointed out that the equations for
form-factors in the $SU(2)$
Thirring model \ref\kirsm{A. N. Kirillov and F. A. Smirnov, Phys. Lett.
198B (1987) 506.}, \ref\smtwo{F. A. Smirnov, Form factors in completely
integrable models of quantum field theory, Advanced Series in
Mathematical Physics 14, World Scientific (Singapure) 1992.}
coincides with the
quantum Knizhnik-Zamolodchikov equation. I. Frenkel and
N. Reshetikhin had investigated in details
the mathematical aspects of quantum
Knizhnik-Zamolodchikov equation and
it's relation to Smirnov's type of form-factors.
 Moreover, in
the recent papers from Kyoto school \japone,\japtwo
\japthree, the
anti-ferroelectric XXZ-Hamiltonian  was diagonalized
in the thermodynamic
limit using the representation theory of quantum affine algebra
$U_q(\widehat{sl(2)})$ .

  In this paper, we address the question that naturally
emerges from the above discussion:
could integrable massive theories be formulated in a similar
fashion as the CFT? This question seems very complicated, so we should
first consider the classical case.

1. Let us recall a few facts about classical phase space
of chiral WZW model \as,\fad,\lf,\gaw,\god.
Consider the group of mappings of interval $[0,2\pi]$
into the finite dimensional
Lie group $SL(2)$ with elements $g(\sigma)$,
 $\sigma \in [0,2\pi]$. The algebra of functions on
 this manifold has natural Poisson structure,
which can be described in the
following way.\foot
{We prefare to speak about Poisson structure
instead of symplectic one.} We denote the Cartan-Weyl basis of
generators of the Lie
algebra $sl(2)$ as $E_+ ,H ,E_-$ . Then the Poisson
brackets for matrix elements $g(\sigma)$
have the form

\eqn\pos{\{g(\sigma)\stackreb{^{\hbox{,}}}
{\otimes}g(\sigma')\}=
(r_{+}\Theta(\sigma-\sigma')+r_{-}\Theta(\sigma'-\sigma))
g(\sigma){\otimes}
g(\sigma'),}
where
\eqn\rmat{r_{\pm}={\pm}h(\frac{H{\otimes}H}{2}+
E_{\pm}{\otimes}E_{\mp})}
and
$$\Theta(\sigma)=
\left\{{1,\ {\sigma}>0 \atop 0,\ {\sigma}<0.}\right.$$
This structure induces linear bracket for currents
 $J(\sigma)=-\frac{2\pi}{h}g^{-1}(\sigma)\ddx g(\sigma)$:
\eqn\km{\frac{1}{2\pi}
\{J^a(\sigma),J^b(\sigma')\}=
f^{abc}J^c(\sigma)
\delta(\sigma-\sigma')-\frac{2\pi}{h}
\delta^{ab}\delta'(\sigma-\sigma').}
In CFT the natural boundary conditions for currents
are periodical ones.
So one can decompose the currents in Fourier series
$$J^a(\sigma)=\sum_{-\infty}^{+\infty}
J_{-n}^ae^{in\sigma},$$
then the commutation relations for generators
$J^a_n$ will have the form        \eqn\mod{\{J^a_n,J^b_m\}=f^{abc}J^c_{n+m}+
\frac{2i\pi}{h}n\delta^{ab}\delta_{n,-m}.}
   After quantization relation \pos\ transforms to
\eqn\cur{g(\sigma)\otimes g(\sigma')=g(\sigma')
\otimes g(\sigma)R,
 \ \  \sigma>\sigma';}
  where $R$ is the universal quantum  $R$-matrix for $U_q(sl(2))$
\ref\dr{V. G. Drinfeld, Quantum groups,
 Proc. ICM-86 (Berkeley), Vol.1
 798  (1987).}
and  \mod\ becomes the commutation relations
for affine Lie algebra.
Let us note that the matrix elements of
quantum field $g(\sigma)$
 are the vertex
operators.
 It is known that the Darboux variables for Poisson structure
\pos\ are classical
analog of Wakimoto free fields. More precisely, in
the Gauss parameterization

\eqn\gauss{\eqalign{g(\sigma)=
\pmatrix{1&0\cr \psi&1\cr}
\pmatrix{e^{\phi/\sqrt2}&0\cr 0&e^{-\phi/\sqrt2}\cr}
\pmatrix{1&\gamma\cr 0&1\cr},}}

$$\psi (\sigma)=\int_{\sigma}^{2\pi}{d\mu} \beta(\mu)e^{-\sqrt2\phi(\mu)}$$
the Poisson structure \pos\ is canonical
\eqn\pstr{\eqalign{&\{\phi(\sigma),\phi(\sigma')\}=
\frac{h}{2}\epsilon(\sigma-\sigma'),\cr
&\{\gamma(\sigma),
\beta(\sigma')\}=h\delta(\sigma-\sigma'),\cr
&\{\phi(\sigma),\gamma(\sigma')\}=
\{\phi(\sigma),\beta(\sigma')\}=\cr
&=\{\gamma(\sigma),\gamma(\sigma')\}=
\{\beta(\sigma),\beta(\sigma')\}=0,}}
where $\epsilon(\sigma)=\Theta(\sigma)-\Theta(-\sigma).$
One can check that  the function
$\int^{2\pi}_0{d\mu}\beta(\mu)e^
{-\sqrt2\phi}$               commutes with whole
current algebra; this object after
quantization is  the screening
operator in Feigin-Fuchs-Dotsenko-Fateev
integral representation for correlation
functions.

This is the classical picture which is the underline of the
present technique in CFT;
although the main technical results could be
obtained in different fashion
(and in many cases they were) we think this
interpretation deserves attention.

2. In this letter the natural, from our point
of view, generalization
of the picture described above,
will be suggested.
This construction is based on  Darboux
variables emerging from Gauss
decomposition.

 Let us consider the group with elements being the
mappings $Z(\theta)\in{SL(2)}$. Keeping in mind the applications of
this construction for massive integrable
models of field theory \sm, \fr, \japone, \japtwo, \japthree,
 we will consider the parameter  $\theta$ as the rapidity,
so the whole line is a natural range of its variation.
 One  should  note this important difference with the
conformal case. The group element
$Z(\theta)$, as $g(\sigma)$,
 admits the Gauss decomposition \gauss. We will preserve the
notation \gauss\ for the components of $Z(\theta)$,
but will consider a
more general form for function $\psi$:
$$Z(\theta)=\pmatrix{e^{\phi/\sqrt2}&\gamma
e^{\phi/\sqrt2}\cr \psi
e^{\phi/\sqrt2}&e^{-\phi/\sqrt2}+e^{\phi/\sqrt2}
\gamma \psi},$$
\eqn\modifi{\psi(\theta)=
\int^{+\infty}_{-\infty}{d\theta'}
\beta(\theta')e^{-\sqrt2\phi(\theta')}
h(\theta'-\theta),}
where kernel $h(\theta)$ is a numerical function.
Moreover, we admit more general form for Poisson
structure for $\phi$,
keeping others
$\beta,\gamma $ the same as in \pstr:
\eqn\nebr{\eqalign{&\{\phi(\theta),\phi(\theta')\}
=\frac{h}{2}\rho(\theta-\theta'),\cr
&\{\gamma(\theta),\beta(\theta')\}=
h\delta(\theta-\theta'),\cr
&\{\phi(\theta),\gamma(\theta')\}
=\{\phi(\theta),\beta(\theta')\}=\cr
&=\{\gamma(\theta),\gamma(\theta')\}=
\{\beta(\theta),\beta(\theta')\}=0,}}
here  $\rho(\theta)$  should be an odd function
\eqn\ro{\rho(\theta)=-\rho(-\theta).}

{}From the requirement that matrix elements of $Z(\theta)$ form quadratic
Poisson algebra under \nebr\ we immediately obtain two functional
equations:

\eqn\main{\rho(\theta)h(\theta')=2h(\theta'-\theta)h(\theta)
-\rho(\theta'-\theta)h(\theta'),}
\eqn\maintwo{\eqalign{&\rho(\mu-\mu')[h(\mu-\theta)
h(\mu'-\theta')-h(\mu-\theta')h(\mu'-\theta)]
=\cr
&=\rho(\theta-\theta')[h(\mu-\theta)h(\mu'-\theta')+h(\mu'-\theta)h
(\mu-\theta')]-\cr
&-2h(\theta-\theta')h(\mu-\theta)h(\mu'-\theta)+2h(\theta'-\theta)
h(\mu-\theta')h(\mu'-\theta').}}

At the same time one could check that the Poisson algebra for
matrix $Z(\theta)$ should have
the form :
\eqn\alg{\{Z(\theta)\stackreb{^{\hbox{,}}}{\otimes}Z(\theta')\}=
r(\theta-\theta')Z(\theta)\otimes Z(\theta'),}
where
\eqn\rmatr{r(\theta)=h[\frac{\rho(\theta)}{2}H\otimes H+h(\theta)
E_{+}\otimes E_{-}-h(-\theta)E_{-}\otimes E_{+}].}
One must  note that the first equation \main, is
nothing more but classical
Yang-Baxter equation for $r$-matrix \rmatr, which is
equivalent to Jacobi
identity for Poisson structure \alg . The nature of
the second
identity \maintwo, seems to be
less clear.

 The systems of functional equations \main,\maintwo\ admits
three type of
solutions \foot{up to equivalence which in terms of
$r(\theta)$ translates as
$$r(\theta-\theta')\longrightarrow\Omega(\theta)\otimes \Omega(\theta')
r(\theta-
\theta')\Omega(\theta)^{-1}\otimes \Omega(\theta')^{-1},
 \Omega(\theta)=\exp(\xi H\theta),$$
where $\xi$ is constant number.}:

\noindent
i. constant solution
\eqn\solone{\rho(\theta)=\epsilon(\theta),h(\theta)=\Theta(\theta);}
ii. rational solution
\eqn\soltwo{\rho(\theta)=2h(\theta)=V.P.\frac{1}{\theta};}
iii. trigonometric solution
\eqn\tre{\eqalign{&\rho(\theta)=V.P.\cth(\kappa \theta), \cr
&h(\theta)=V.P.\frac{1}{1-exp(-2\kappa \theta)};}}
where $\kappa$  is an arbitrary parameter and we denote by V.P. the
principal value.

All of this solution are well known in theory of
classical Yang-Baxter
equation \ref\bd{A. A. Belavin and V. G. Drinfeld,
Funct. Anal. i ego Pril. 16 (1982) 1 (in  Russian).}.
The trigonometric solution is the
most general one; the constant and rational solutions
could be obtained
from it in the limits $\kappa\longrightarrow +\infty$ and
$h,\kappa\longrightarrow 0 ; \frac{h}{\kappa}\longrightarrow const$
respectively.
Also, for the limit $\kappa \longrightarrow \infty$
$Z(\theta)$ obeys the same algebra as $g(\sigma)$,
$\psi$ from \modifi\ has the form \gauss\
 and the Poisson structure
\nebr\ tends to (7). Thus,  bellow we'll consider only
trigonometric solution.

  We will define the currents in the spirit of the work \sm. In fact
two type of currents could be defined:
\eqn\fedd{L(\theta)=Z(\theta+\frac{1}{\kappa})Z^{-1}(\theta),}
\eqn\fed{L'(\theta)=Z^{-1}(\theta)Z(\theta+\frac{1}{\kappa}).}
It is simple to check that in the limit $\kappa\longrightarrow +\infty$
the currents  $L'(\theta)$ can be written in the
 form $L'(\theta)=1-\frac{2\pi}{h\kappa}J(\theta)+O(\frac{1}{\kappa^2})$,
 where
\eqn\wak{\eqalign{&-\frac{2\pi}{h}J_3=\ddt\phi+{\sqrt2}\beta\gamma, \cr
&-\frac{2\pi}{h}J_-=-\beta ,\cr
&-\frac{2\pi}{h}J_+=\beta{\gamma}^2+\sqrt2\gamma{\ddt\phi}+\ddt\gamma}}
and the Poisson brackets for  variables  $\gamma,\beta,\phi$
are defined by \pstr.
One can recognize in \wak\ the classical limit of
the Wakimoto
representation for currents, so  currents in \wak\
satisfy the Poisson            algebra \km.

The currents  $L(\theta)$ from \fedd\ form the quadratic algebra  :
\eqn\rsts{\eqalign{&\{L(\theta)\stackreb{^{\hbox{,}}}{\otimes}
L(\theta')\}=r(\theta-\theta')L(\theta){\otimes}L(\theta')+
L(\theta){\otimes}L(\theta')
r(\theta-\theta')\cr
&-L(\theta){\otimes}1
r(\theta-\theta'-\frac{1}{\kappa})1
{\otimes}L(\theta')-
1{\otimes}L(\theta')
r(\theta-\theta'+\frac{1}{\kappa})L(\theta){\otimes}1,}}
the latter  is the algebra defined by Reshetikhin-Semenov-Tian-Shansky
\ref\rst{N. Yu. Reshetikhin and M. A. Semenov-Tian-Shansky,
Lett. Math. Phys. 19 (1990) 133.}. They had shown in \rst\
that \rsts\ is the classical limit of
quantum affine algebra $U_q(\widehat{sl(2)})$  \dr.
The limiting procedure $\kappa \longrightarrow \infty$
,when $Z \longrightarrow g$ differs from one
described in \rst\ for $L(\theta)$.
The procedure
of \rst\ requires the replacement  $\kappa \longrightarrow i\kappa$
in  \fedd and \rsts. We must note that there are difficulties in this
analytical continuation of our expression for $Z(\theta)$, because
 of the presence
of principal value in its definition ( see \modifi,\nebr,\tre ).

  Allowed boundary conditions for the fields $L(\theta)$,
 as well as natural expansion
for them, are those that are consistent with the kernel \tre\ and
Poisson structure \nebr. From the other side what are
natural boundary conditions  from the point of view
of massive
Field Theory for $Z(\theta)$ and $L(\theta)$ is not  clear at all.
It seems to be related to the question  of
analytical properties, mentioned above.
This is important to understand
for further progress. We hope that the free field
representation may help to make it clear.

  At the end, let us  recall that the quantum
operators $Z(\theta)$ in
massive integrable theories have a meaning of
creation operators for
asymptotic states, i.e. some kind of Faddeev-Zamolodchikov
operators.
One would like to have
the proper quantum analog of the free field representation
for $Z(\theta)$
which will allow
to construct integral representation for correlation
functions in integrable  models.

  Note added. When this work was finished we received
the papers \ref\japnew1{Jun'ichi Shriaishi,
Free boson representation of $U_q(\widehat{sl(2)})$,
Tokyo preprint, UT-617, september 92.},
\ref\japnew2{Akishsi Kato, Yas-Hiro Quano and
Jun'ichi Shriaishi, Free boson representation
of q-vertex operators and their
correlation functions, Tokyo preprint UT-618,
september 92.}, where
the free field representation of
$U_q(\widehat{sl(2)})$
in different approach was developed.
It would be interesting to
understand the
relation of these two pictures that
looks very different at this moment.

Acknowledgment: We are grateful to L. D. Faddeev, V. A. Fateev,
I. B. Frenkel, N. Yu. Reshetikhin and A. B. Zamolodchikov
for useful discussions.
Part of this work was done during the visit of one
of us (S.Sh) to
String Theory Group at Rutgers; he thanks the
members of this group for warm hospitality.

\listrefs

\end